\documentstyle[12pt]{article}   

\large
\newtheorem{definition}{Definition}[section]

\makeatletter
\@addtoreset{equation}{section}
\makeatother

\newcommand{\beq}{\begin{equation}}
\newcommand{\eeq}{\end{equation}}
\newcommand{\be}{\begin{equation}}
\newcommand{\ee}{\end{equation}}
\newcommand{\ba}{\begin{eqnarray}}
\newcommand{\ea}{\end{eqnarray}}
\newcommand{\AmG}{{\cal A}/{\cal G}}
\newcommand{\AmGb}{\overline{\AmG}}
\newcommand{\Lp}{{\cal L}_p}
\newcommand{\HG}{{\cal HG}}
\newcommand{\HA}{{\cal HA}}

\newtheorem{Lemma}{Lemma}[section]

\def\be{\begin{equation}}
\def\ee{\end{equation}}
\def\ba{\begin{eqnarray}}
\def\ea{\end{eqnarray}}

\def\ag{{{\cal A}/{\cal G}}}

\def\agb{{\overline {{\cal A}/{\cal G}}}}

\def\Comp{{\mathchoice
{\setbox0=\hbox{$\displaystyle\rm C$}\hbox{\hbox to0pt
{\kern0.4\wd0\vrule height0.9\ht0\hss}\box0}}
{\setbox0=\hbox{$\textstyle\rm C$}\hbox{\hbox to0pt
{\kern0.4\wd0\vrule height0.9\ht0\hss}\box0}}
{\setbox0=\hbox{$\scriptstyle\rm C$}\hbox{\hbox to0pt
{\kern0.4\wd0\vrule height0.9\ht0\hss}\box0}}
{\setbox0=\hbox{$\scriptscriptstyle\rm C$}\hbox{\hbox to0pt
{\kern0.4\wd0\vrule height0.9\ht0\hss}\box0}}}}
\def\Co{{\mathchoice
{\setbox0=\hbox{$\displaystyle\rm C$}\hbox{\hbox to0pt
{\kern0.4\wd0\vrule height0.9\ht0\hss}\box0}}
{\setbox0=\hbox{$\textstyle\rm C$}\hbox{\hbox to0pt
{\kern0.4\wd0\vrule height0.9\ht0\hss}\box0}}
{\setbox0=\hbox{$\scriptstyle\rm C$}\hbox{\hbox to0pt
{\kern0.4\wd0\vrule height0.9\ht0\hss}\box0}}
{\setbox0=\hbox{$\scriptscriptstyle\rm C$}\hbox{\hbox to0pt
{\kern0.4\wd0\vrule height0.9\ht0\hss}\box0}}}}
\def\Rl{{\mathchoice
{\setbox0=\hbox{$\displaystyle\rm R$}\hbox{\hbox to0pt
{\kern0.4\wd0\vrule height0.9\ht0\hss}\box0}}
{\setbox0=\hbox{$\textstyle\rm R$}\hbox{\hbox to0pt
{\kern0.4\wd0\vrule height0.9\ht0\hss}\box0}}
{\setbox0=\hbox{$\scriptstyle\rm R$}\hbox{\hbox to0pt
{\kern0.4\wd0\vrule height0.9\ht0\hss}\box0}}
{\setbox0=\hbox{$\scriptscriptstyle\rm R$}\hbox{\hbox to0pt
{\kern0.4\wd0\vrule height0.9\ht0\hss}\box0}}}}

\title{An axiomatic approach to Quantum Gauge Field Theory}
\author{T. Thiemann\thanks{New address: Physics Department, Harvard
University, Cambridge, MA 02138, USA, Internet:
thiemann@harvard.math.edu},\\ Physics Department, The
Pennsylvania State University,\\ University Park, PA 16802-6300, USA}
\date{{\small Preprint CGPG-95/7-1, Preprint HUTMP/B-345}}

\begin{document}

\maketitle                     

\begin{abstract}

In the present article we display a new constructive quantum field theory
approach to quantum gauge field theory, utilizing the recent progress
in the integration theory on the moduli space of generalized connections
modulo gauge transformations.\\
That is, we propose a new set of Osterwalder Schrader like
axioms for the characteristic
functional of a measure on the space of generalized connections modulo gauge
transformations rather than for the associated Schwinger distributions.\\
We show non-triviality of our axioms by demonstrating that they are satisfied
for two-dimensional Yang-Mills theory on the plane and the cylinder.\\
As a side result we derive a closed and analytical expression
for the vacuum expectation value of an arbitrary product of Wilson loop
functionals from which we derive the quantum theory along the Glimm and Jaffe
algorithm which agrees exactly with the one as obtained by canonical methods.

\end{abstract}

\section{Introduction}

It is an old dream of theoretical physicists to base the description of
Yang-Mills (gauge) theories on the so-called Wilson-Loop observables. These
are simply (products of) traces of holonomies around closed loops in the
given spacetime. Also general relativity, when formulated as a dynamical
theory of connections, can be explored via Wilson loop variables \cite{15}.\\
The advantage of these observables is that they provide for an overcomplete
set of coordinates for the gauge invariant information that is contained in
the connection \cite{5}, also called the moduli space of (spacetime)
connections modulo
gauge transformations, $\AmG$. There are two major disadvantages :\\
1) The space $\AmG$ is nonlinear. Therefore, all the mathematical physics
techniques that have been developed for field theories whose underlying
space of fields is linear are not available. A solution out of this is to
fix a gauge and to work with Schwinger functions of connections in that
gauge but that comes at a price : manifest gauge invariance is lost and we
have the problem of the annoying Gribov copies. Also from a geometrical
viewpoint it just does not seem to be right to enforce linearity by brute
force.\\
2) If one keeps manifest gauge invariance, and the only way to do this as far
as we know is to work with Wilson-Loop functionals\footnote{for instance,
invariants constructed from the curvature suffer from the field copy
problem \cite{24}}, then the the connection
is smeared with a loop function. There is then an immediate problem : it is
well-known that linear quantum fields are rather distributional and need to
be smeared {\em in all spacetime directions}, however, a loop only smears in
one such direction. That either means that we have to give up this approach
or that YM quantum fields are simply better behaved in the precise sense that
the (vacuum expectation value of the) Wilson observables exists. There is a
chance that this is true, at least in the non-Abelian case, since the
physically relevant phase of, say, QCD is not described by the Fock
Hilbert space.\\
In this article we take advantage of the existence of new integration
techniques developed in \cite{2,3,4} in order
to set up a system of Osterwalder-Schrader (OS) axioms
that are tailored to $\AmG$ \cite{26}. Our axioms are imposed directly on
the measure rather than on the associated Schwinger distributions \cite{13}
and is thus more in the fashion of \cite{11}. This will enable us to
circumvent all the problems that are connected with these earlier approaches.
Our approach is as rigorous the the ones in \cite{11} for the
linear case or \cite{13} for the YM case.
Furthermore, we prove non-triviality of these axioms by showing that
they have a non-trivial solution, namely we verify them for two-dimensional
pure YM theory for any semi-simple compact gauge
group which is known to be an integrable, finite dimensional model. \\
\\
The paper is organized as follows :

In section 2 we review the relevant notions from calculus on $\AmG$.

In section 3 we motivate and introduce a new set of axioms tailored
to quantum gauge field theory.

In section 4 we derive the general form for the generating functional of the
Yang-Mills measure on $\AmG$ for any compact semi-simple gauge group for
the two-dimensional spacetimes of the topology of the plane and
the cylinder which are the ones of physical relevance.

In section 5 we explicitly verify the new axioms for the model analyzed
in section 4 and give the relation to the Hamiltonian approach.

\section{Preliminaries}

\subsection{Integration on $\AmG$}

We review here the necessary notions from \cite{2,3,4,26} and references
therein.\\
We will consider the set $\Lp$ of oriented unparametrized loops based at an
arbitrary but fixed point $p$ of spacetime $M$ as the entity of
piecewise analytical embeddings of the circle into $M$. Throughout this
paper we will deal only with based, piecewise analytical loops. With
respect
to the natural composition of loops, $\Lp$ adopts the structure of a
semi-group. With the help of the space $\AmG$ of smooth connections modulo
gauge transformations we turn $\Lp$ into a group $\HG$, called the hoop
group in the sequel, as follows :\\
We define two loops $\alpha_1,\alpha_2\in\Lp$ to be holonomically
equivalent, $\alpha_1\sim\alpha_2$, iff their holonomies agree on every
point $A\in\AmG$, that is,
$h_{\alpha_1}(A)=h_{\alpha_2}(A)\forall A\in\AmG$. Here the holonomy
map is defined as the path-ordered exponential of the line integral
of the connection along the loop :
\beq h_\alpha(A):={\cal P} \exp(\oint_\alpha A) \;. \eeq
The symbol $\cal P$ asks that in any parameterization of the loop the
terms with the highest values of the parameter be ordered to the left.
Then $\HG:=\Lp/\sim$. We will not distinguish any more in the sequel between
a hoop and its various representants and will use the word loop again
unless confusion could arise.\\
We will assume once and for all that the gauge group $G$ is compact and
semisimple of rank r (the non-semisimple case can be treated in a similar
manner). The associated principal fibre bundle is taken to be trivial.\\
We now introduce the so-called Wilson-loop functionals \cite{5}
\beq T_\alpha(A):=\frac{1}{N}\mbox{tr}(h_\alpha(A)) \eeq
where the trace is taken with respect to the N-dimensional fundamental
representation of $G$.
The Wilson-loops are manifestly gauge invariant functions on
$\AmG$ and are separating on $\AmG$ in the sense that given all
the $T_\alpha$, we can reconstruct the smooth connection modulo gauge
transformations \cite{5}.\\
These quantities enable us to construct an Abelian $C^*$ algebra
as follows : consider the quantities of the form
\[ \Psi:=\sum_{I=1}^r\sum_{i=1}^n z_{Ii} \prod_{J=1}^I
T_{\alpha_{Ii,J}}\;. \]
The system of these objects is easily checked to be an Abelian algebra : the
Mandelstam identities \cite{5} reveal that every product of traces of
the holonomy can be written as a linear combination of products of Wilson
loops with at most $r$ factors. Moreover, it is an Abelian  $^*$
algebra since $\bar{T}_\alpha=T_{\alpha^{-1}}$. Finally, we turn it into
an Abelian $C^*$ algebra by completing it with respect to the norm
\beq ||\Psi||:=\sup_{A\in\AmG} |\Psi|(A) \; . \eeq
The $C^*$ property follows easily from that for complex numbers.\\
This Abelian $C^*$ algebra will be called the holonomy algebra $\HA$.\\
We can now employ usual Gel'fand theory : The Gel'fand spectrum $\AmGb$
of generalized connections is in one-to-one correspondence with the space of
{\em all} homomorphisms
from $\HG$ into the gauge group $G$, that is to say, it is the algebraic
dual of $\HG$. By the Riesz-Markov theorem, regular Borel measures on
$\AmGb$ are in
one-to-one correspondence with positive linear functionals on the space
of continuous functions on $\AmGb$ (recall that with respect to the
Gel'fand topology $\AmGb$ is a compact Hausdorff space).\\
An interesting example of a measure $\mu_0$ on $\AmGb$ has been
constructed in \cite{2} :\\
Consider the family of all piecewise analytical oriented graphs $\Gamma$ in
$M$, that is, piecewise analytic embeddings of closed intervals in $M$.
We will denote its fundamental group by $\pi_1(\Gamma)$. Choose
a system of generators $\beta_1,..,\beta_n$ of $\Gamma$ where
$n:=d_\Gamma:=\dim(\pi_1(\Gamma))$ is the dimension of the fundamental
group. A cylindrical function f on $\AmGb$ can be written as the pullback
under the following map for one of the graphs $\Gamma$ :
\beq p_\Gamma\; :\; \AmGb\rightarrow G^{d_\Gamma}\; ;\; A\rightarrow
(h_{\beta_1}(A),..,h_{\beta_{d_\Gamma}}(A)) \; , \eeq
that is, $f=(p_\Gamma)^* f_\Gamma\mbox{ where }f_\Gamma$ is a map from
$G^{d_\Gamma}$ into the complex numbers.\\
The measure $\mu_0$ is then defined to be the following linear functional
for cylindrical functions
\beq \int_{\AmGb} d\mu_0(A) f(A):=\int_{\AmGb}
d\mu_\Gamma(A) f_\Gamma(p_\Gamma(A)):=\int_{G^{d_\Gamma}}d\mu_H(g_1)..
d\mu_H(g_{d_\Gamma}) f_\Gamma(g_1,..,g_{d_\Gamma}) \; . \eeq
That this defines indeed an infinite dimensional ($\sigma$-additive) measure
$\mu_0$ as the projective limit \cite{12} of the measures $\mu_\Gamma$
defined in (2.5) was shown in \cite{4}.\\
The rigorously defined measure $\mu_0$ will be used in the next section
to construct the Yang-Mills measure. \\
Here are two more definitions which prove useful in the sequel.
\begin{definition}
A loop network state on a given graph $\Gamma$ with fundamental group
$[\beta_1,..,\beta_n]$ is labelled by a triple
$(\Gamma,\vec{\pi},c)$ consisting of that graph $\Gamma$, a vector of
non-trivial irreducible representations $\vec{\pi}=[\pi_1,..,\pi_n]$
and a contraction matrix $c$ which takes values in the projectors onto
the orthogonal irreducible representations contained in the decomposition
of $\otimes_{k=1}^n \pi_k$. It is defined by
\be \label{2.1}
T_{\Gamma,\vec{\pi},c}(A):=\mbox{tr}[\otimes_{k=1}^n\pi_k(h_{\beta_k}(A))
\cdot c] \;. \ee
\end{definition}
Different choices of generators of $\pi_1(\Gamma)$ lead to unitarily
equivalent bases.\\
The loop-network states can be seen to provide for a complete (and
orthogonal with respect to $\mu_0$) basis of states
for any cylindrical subspace of $C(\agb)$ defined by a graph $\Gamma$
\cite{20,21}. In particular, the multiloop states
$T_{\alpha_1}..T_{\alpha_r}$
can always be expressed in terms of those so that we arrive at the
following definition :
\begin{definition}
The characteristic functional of a measure on $\agb$ is
defined by the set of expectation values of loop network states :
\be \label{2.2}
\chi_\mu(\Gamma,\vec{\pi},c):=\int_\agb d\mu(A) T_{\Gamma,\vec{\pi},c}(A)\;.
\ee
\end{definition}

\section{A proposal for Constructive Quantum Gauge Field Theory}

As already mentioned before, the concepts introduced in the textbook
treatments \cite{11} of constructive scalar field theory  seem inadequate
for non-linear theories such as gauge field theories whose space of
histories is given by $\AmG$. The idea is to come up with new axioms that
are guided by the ones for scalar field theory but take the nonlinearity
of $\agb$ fully into account.\\
Although this idea is not completely new, related contributions \cite{13}
seem to be too strongly attached to techniques applicable to linear theories,
mainly because these works are based on axioms for Schwinger distributions.
{}From our point of view it seems much more natural to impose the axioms on
the underlying measure.\\
The OS axioms for a QFT based on a linear space of histories can be roughly
described as follows \cite{11} :\\
One states everything in terms of the characteristic functional $\chi$
of a measure $\mu$ (its Fourier transform) which is required to be
continuous and positive definite on (finite subspaces of) the space
$\cal S$ of test functions of rapid decrease
\beq \chi(f):=<\exp(i\Phi[f])>:=\int_{{\cal S}'}d\mu(\Phi)
\exp(i\Phi[f]) \eeq
and $\Phi[f]:=\int_{R^{d+1}} d^{d+1}x \Phi(x)
f(x)$ denotes the canonical pairing between distributions and test
functions.\\
As this $\cal S$ is a nuclear space, Minlos' theorem \cite{12} then tells
us that (as already displayed in (3.1))
the measure has
support on the space ${\cal S}'$ of tempered distributions. It is obvious
that right from the beginning everything is soaked into kinematical
linearity.\\
In order to find the appropriate analogue of these notions for constructive
gauge field theory, let us make some heuristic considerations :\\
The counterpart of the scalar field $\Phi$ is of course the connection.
Since we are interested in a measure theoretic formulation of the
theory, we now have to look for the analogue of the expression
$\exp(i\Phi[f])$. Let us look for a moment at the Abelian case.
Then the Wilson loop functional is given by
$T_\alpha=\exp(iA[\alpha])$
where we have allowed for a distributional connection and the
canonical pairing between the field $A$ and the loop $\alpha$ is now
given by $A[\alpha]=\oint_\alpha A=\int_0^1 dt \dot{\alpha}^a
A_a(\alpha(t))$ rather than $\Phi[f]=\int d^{d+1}x\Phi(x)
f(x)$.
This is an important difference : in order that this object makes sense,
the connection is not allowed to be in ${\cal S}'(\Rl^d)$ ! This
immediately
implies that the theory that we want to base on Wilson loops will not
result in the usual Fock space, not even for Maxwell theory !\\
However, the formal similarity between the expressions $\exp(i\Phi[f])$ and
$T_\alpha$ generalized to the non-Abelian case thus motivates to base
the generating functional $\chi$ of a measure on $\AmGb$ on the
usage of Wilson loops. Since for a rank $r$ group products of
Wilson loop functionals can only be reduced to sums of products of
at most r Wilson loop functionals, we arrive naturally at
the expression (2.7) for the characteristic functional of a
measure on $\agb$. Accordingly, the analogue of the probes of the
the field $\Phi$, namely test functions of rapid decrease, are piecewise
analytic loops in Euclidean space.\\
The nice thing is that the precise analogue of Bochner's theorem can be
argued to be the Riesz-Markov theorem : any positive linear functional $\chi$
on $C(\agb)$ gives rise to a regular Borel measure $\mu$ on $\agb$ which
is a compact Hausdorff space by construction.\\
What we do not have is an analogue of the Minlos theorem which is due
to the fact that we did not specify any topology on the
space of probes, i.e. the set of loops
(see \cite{22} for an attempt to build a nuclear topology on
$\cal HG$).
Since we will never need to specify the analogue of Schwinger distributions
in what follows, we do not worry about that. Suffice it to say that
our carrier space $\agb$ is the maximal extension of $\ag$ such that Wilson
loops are still continuous (with respect to the Gel'fand topology on
$\agb$) and we do not mind working with that bigger space although the
actual carrier of physically relevant measures maybe significantly smaller.\\
\\
Let us now formulate analogues of the OS axioms \cite{11} :\\
\\
{\bf A Quantum gauge field theory is a probability measure $\chi$ on
$\agb$ satisfying the following axioms} :\\
$\bullet$ OS-I) Analyticity\\
This axiom in scalar QFT ensures that Schwinger functions of all orders
exist. Since we are not interested in these for gauge field theory,
because they are not gauge invariant, we will drop that axiom here
altogether ! \\
$\bullet$ OS-II) Regularity\\
The regularity axiom for the scalar field prescribes some bound on
the characteristic functional. Technically, it can be used to show that
the measure is supported on the space of tempered distributions, rather than
on those which are continuous on the test functions of compact support.
This is important if we want to do things like Wick rotations
of Schwinger functions to Wightman functions. Since we are not interested
in that issue, we also simply drop this axiom !\\
$\bullet$ OS-III) Euclidean invariance\\
The action $g\cdot\Gamma$ of an element $g$ of the full Euclidean E group
in d spacetime dimensions on a graph is just the image of the linear
transformation $x\to(g\cdot x)$ where $x$ is a point on $\Gamma$.
The measure is required to be invariant under
this action \[ \chi(g\cdot\Gamma,\vec{\pi},c)=\chi(\Gamma,\vec{\pi},c). \]
$\bullet$ OS-IV) Reflection positivity\\
This is the most important of the axioms because it allows to reformulate the
theory in terms of more familiar concepts, that is, it provides us with a
notion of time, a Hilbert space, and a Hamiltonian (compare \cite{11} for
the proof of this fact which is completely {\em insensitive} to whether the
space of histories is linear or not\footnote{This observation is due to
Abhay Ashtekar}).
The technicalities are as follows :\\
Choose an arbitrary hyperplane in $\Rl^d$ which we will call the
time zero plane. Consider the linear span, denoted V, of the following
functions on $\AmGb$ of the form
\[ \Psi_{\{z_{I}\}} \; :\; \AmGb\to \Co\; ; A\to\sum_{I\in S} z_I T_I \]
where each index $I$ stands for data $I=(\Gamma,\vec{\pi},c)$ of a loop
network and where $\Gamma$ is supported in the positive time half space
$\{x=(x^0,\vec{x})\in R^d\; ;\; x^0 >0\}$ and $S$ is a finite set
of indices.
Furthermore, let $\Theta(x^0,\vec{x})=(-x^0,\vec{x})$ denote the time
reflection operator ($\Theta\in E$).
Then it is required that for each $\psi,\Xi\in V$
\beq (\Psi,\Xi):=<\Theta\Psi,\Xi>:=\int_\agb d\mu(A)
\overline{\Theta\Psi[A]}\Xi[A] \ge 0 \eeq
is non-negative. \\
Reflection positivity has already been verified on the (finite) lattice
for YM theory \cite{11}.\\
$\bullet$ OS-V) Ergodicity\\
This axiom ensures the uniqueness of the vacuum (a vector invariant under
the time translation subgroup of E, $T(s)(x^0,\vec{x})=(x^0+s,\vec{x})$). The
requirement is that
\be \lim_{t\to\infty} \frac{1}{t}\int_0^t ds (T(s)\Psi)[A_0]
=\int_\agb d\mu(A)\Psi[A] \ee
for any vector $\Psi\in L_1(\agb,d\mu)\cap V\mbox{ and any }A_0\in
\agb$. Note that the rhs implies that the lhs does not depend on the
particular choice of $A_0$.\\

\section{Euclidean YM gauge theory in two dimensions}

\subsection{The lattice regularization}

Consider a special family of graphs $\Gamma(a,L_x,L_y)$ in $M$, namely
finite square lattices with spacing $a$ of length $L_x
\mbox{ and } L_y$ in x and y direction respectively with respect to the
Euclidean norm of the 2-dimensional Euclidean spacetime M (i.e. the
spacetime metric is $g_{ab}=\delta_{ab})$.
Thus we have introduced an IR regulator (the finite volume defined by the
$L_x\mbox{ and }L_y$) and an UV regulator (defined by the lattice
spacing $a$). We
have $(N_x+1)(N_y+1)$ vertices on that finite
lattice where $N_x a:=L_x,\; N_y a:=L_y$.\\
An open path along an edge (link) l of the lattice from
the vertex i to the vertex j will be denoted by
\beq l=l_{i\to j} \eeq
which enables us to define the
plaquette loops $\Box$ based at $(x,y)$ according to
\beq \Box_{(x,y)}:=l_{(x,y)\to(x,y+1)}^{-1}\circ
l_{(x,y+1)\to(x+1,y+1)}^{-1}\circ
l_{(x+1,y)\to(x+1,y+1)}\circ l_{(x,y)\to(x+1,y)}. \eeq
That is, the plaquette loop starts at the bottom left corner and our
convention is such that the coordinate directions define positive
orientation. Here the coordinates $x,y$ are taken to be integers (in
lattice units).\\
There are no boundary conditions for the plane $M=R^1\times R^1$ while we
identify\newline
$l_{(1,y)\to(1,y+1)}\mbox{ and } l_{(N_x+1,y)\to(N_x+1,y+1)}$
on the cylinder $M=R^1\times S^1$.\\
We will choose the basepoint $p$ to lie in the upper right corner of
$\Gamma$ and we will use the following generators of $\pi_1(\Gamma)$ :\\
1) On the plane, choose an open path $\rho_{x,y}$ within $\Gamma$ from
$p$ to the point $(x,y)$. Then we have the $N_x N_y$ generators
\beq \beta_{x,y}:=\beta_{\Box_{(x,y)}}:=\rho_{x,y}^{-1}\circ\Box_{(x,y)}\circ
\rho_{x,y} \; . \eeq
2) On the cylinder we need apart from (4.3) one more generator ``which
wraps once around the cylinder". We will choose the horizontal loop
``at future time infinity"
\beq \beta_x:=l_{(N_x,N_y+1)\to(1,N_y+1)}\circ l_{(N_x-1,N_y+1)\to
(N_x,N_y+1)} \circ..\circ l_{(1,N_y+1)\to (2,N_y+1)} \;. \eeq
We are now ready to define the regulated characteristic functional of the
regulated $G$ Yang-Mills measure :\\
Consider the following cylindrical functions on $\AmGb$, cylindrical
relative to our lattice graph $\Gamma$ :\\
1) the exponential
\beq \exp(-\beta S_{Wilson})\mbox{ where }S_{Wilson}(A):=\sum_\Box
[1-\frac{1}{N}\Re \mbox{tr}(h_\Box(A))] \eeq
is the so-called Wilson action for $G$ Yang-Mills theory \cite{8},
$h_\Box$ is the holonomy along the based loop $\Box$, $\Re$tr means
``take the real part of the trace of'' and
the ``inverse temperature" is given by
\beq \beta=\frac{1}{g_0^2 a^{4-d}} \eeq
($d=2$ is the dimension of M) where $g_0=g_0(a)$ is the bare coupling.
Here we
have chosen a basis $\{\tau_I\}_{I=1}^{\mbox{dim}(G)}$ for the Lie algebra
$L(G)$ of $G$ such that tr$(\tau_I\tau_J)=-N\delta_{IJ}$.\\
2) the product of Wilson-loop functionals for any $r$ loops
$\alpha_1,..,\alpha_r$
embedded in $\Gamma=\Gamma(a;L_x,L_y)$ and $N$ is the dimension of the
fundamental representation of $G$.\\
These functions are cylindrical since the loops $\alpha_i$ can, by
definition of the fundamental group of a
graph, be expressed as a particular composition of the generators of
$\pi_1(\Gamma)$.\\
Thus, all these functions are measurable with respect to the measure
$\mu_0$ introduced in section 2 and the following definition
makes sense :
\ba
& &\chi(\alpha_1,..\alpha_{N-1};a;L_x,L_y):=<T_{\alpha_1}..T_{\alpha_r}>
\nonumber\\
&:=& \frac{1}{Z(a;L_x,L_y)}\int_{\AmGb} d\mu_0(A) e^{-\beta
S_{Wilson}(A)}T_{\alpha_1}(A).. T_{\alpha_r}(A) \nonumber\\
&:=&\frac{1}{Z}\prod_{\beta\in\pi_1(\Gamma)}\int_G
d\mu_H(h_\beta) \exp(-\beta S_{Wilson}) \mbox{tr}(\prod_{\beta\in\alpha_1}
h_\beta)..\mbox{tr}(\prod_{\beta\in\alpha_r} h_\beta)
\ea
where the notation $\beta\in\alpha_i$ means ``composition of all those
generators $\beta$ necessary to express $\alpha_i$ (in the specific order as
defined by $\alpha_i$)". \\
The partition function $Z=Z(a;L_x,L_y)$ is defined through
$\chi(p,..,p;a;L_x,L_y)=1$ where $p$ is, as above, the
basepoint of all the loops on the lattice (it is in particular a trivial
loop).\\
The idea is now quite similar to related constructions in constructive
quantum scalar field theory \cite{19}. There, one integrates the regularized
version of the exponential of $-\beta$ times the interaction part of the
Euclidean action times $\exp(i\Phi[f])$ with the rigorously defined free
(Gaussian) measure.
Then one takes the thermodynamic (infinite volume) and continuum limit
of the resulting expression and obtains the characteristic functional of
a rigorously defined interacting theory.\\
In our case the role of the Gaussian measure is played by the
rigorously defined, $\sigma$-additive measure $\mu_0$ on the universal
carrier $\agb$, the
``regularized interaction part of the action'' is played by the Wilson
action and the analogue of $\exp(i\Phi[f])$ is given by the products
of Wilson loop functionals.

The reader might worry that we have changed lattice gauge field theory in
the previous section \cite{8}. One can show that this is not the case, i.e.
both formulations are equivalent \cite{23}.

\subsection{Expression of the Wilson loop in terms of plaquettes}

By definition of the fundamental group of a graph, each of the $\alpha_i$
involved in the characteristic functional $\chi$ can be written as
a particular composition of the generators of the graph (lattice).
In this section we are going to characterize this composition.\\
First three definitions :
\begin{definition}
i) A loop is said to be simple iff there is a holonomically equivalent
loop which has no self-intersections.\\
ii) By the surface enclosed by a simple loop we mean the surface that is
1) bounded by the simple loop and 2) lies to the left as one follows the
loop counterclockwise (mathematically positive direction).\\
iii) Two distinct simple loops are said to be non-overlapping iff the
intersection domain of the surfaces that they enclose have zero Euclidean
area. \end{definition}
So, for example all the loops $\beta_{x,y}$ are simple since they lie
in the same hoop class as the plaquette loops $\Box_{(x,y)}$.
Non-overlapping distinct simple loops may share whole segments whence
the plaquette generators of our graph (lattice) are mutually
non-overlapping.\\
The following two simple lemmas govern the form of the characteristic
functional in two spacetime dimensions.
\begin{Lemma}
Every simple, homotopically trivial loop on $\Gamma$ can be written
as a particular composition of the generators $\beta_\Box$ whose surfaces
are contained in the surface enclosed by that loop, each of them appearing
once and only once.
\end{Lemma}
Proof :\\
We will do this by direct construction.\\
Consider a simple, based loop of the form
$\alpha=\rho^{-1}\circ\beta\circ\rho$ where $\rho$ is the
given open path between $p$ and the starting point $q$ on the unbased
loop $\beta$. Subdivide $\beta$ into columns parallel to the y-axis.
If q does not coincide yet
with the vertex of the horizontal link with the lower x argument on the
bottom of the most right column then let $l$ be the edge of $\beta$ between
$q$ and that point on $\beta$. Then we have
$\alpha=(l\circ\rho)^{-1}\circ(l\circ\beta\circ\l^{-1})\circ(l\circ\rho)$ so
that by appropriate redefinition of $\beta$ and $\rho$
we can always achieve that the starting point $q$ on $\alpha$ is the point
mentioned above.\\
Let $q:=(x,y)$ and $n$ be the height of that most
right column. Then the holonomy around $\alpha$ is given
by ($g$ is the product of the holonomies of those links that are not
involved in the most right column)
\begin{eqnarray}
& & h_\rho^{-1} g h_{(x,y+n)\to(x+1,y+n)}^{-1} h_{(x+1,y+n-1)\to(x+1,y+n)}
...h_{(x+1,y)\to(x+1,y+1)} h_{(x,y)\to(x+1,y)} h_\rho \nonumber\\
&=& \mbox{Ad}[h_\rho^{-1}](g h_{(x,y+n-1)\to(x,y+n)}..h_{(x,y)\to(x,y+1)})
\nonumber\\
& & \mbox{Ad}[(h_{(x,y+n-2)\to(x,y+n-1)}..h_{(x,y)\to(x,y+1)}
h_\rho)^{-1}](\Box_{x,y+n-1})..\mbox{Ad}[h_\rho](\Box_{x,y})\nonumber\\
&=:& \{\mbox{Ad}[h_\rho^{-1}](h_{\tilde{\beta}})\} g_{x,y+n-1}..g_{x,y}
\end{eqnarray}
Here we have denoted by Ad the adjoint action of the group on itself, that
is, $\mbox{Ad}[g](h):=ghg^{-1}$.\\
The curly bracket in the last line is the holonomy around $\alpha$
with the most right column removed and the lost segment, resulting from that
removement, reattached. This loop we called
$\rho^{-1}\circ\tilde{\beta}\circ\rho$. The remaining product involves all
the
based plaquette loops of the most right column, each of them appearing
precisely once. We iterate
like this to the left until we reach the last column.
But the analogous curly bracket term as the one above for the last column is
the identity.\\
$\Box$\\
\begin{Lemma}
Every loop can be written as a composition of simple non-overlapping loops.
\end{Lemma}
Proof :\\
Given a loop $\alpha$ on $\Gamma$, consider the pattern of surfaces on
$M$ that its homotopically trivial part defines (the homotopically
non-trivial loop $\beta_x$ are not overlapping with any other
loop and do not enclose a surface). Take the boundaries of
these surfaces as the
definition of the simple non-overlapping loops $\alpha_I,\;I=1,..,n$,
and note that the pattern of surfaces defines a subgraph $\tilde{\Gamma}$
of $\Gamma$. Since the $\alpha_I$ are simple,
by the preceding lemma, we can express one (call it $\beta_I$) of the
$\beta_{x,y}$ contained in the surface enclosed by $\alpha_I$ by
$\alpha_I$ and the rest of these $\beta_{x,y}$. This is possible because
the corresponding transformation is non-singular (the $\beta_{x,y}$
contained in the surface enclosed by $\alpha_I$ appear precisely once).
Since the $\alpha_I$ are non-overlapping, we conclude that the $\beta_I$
are all distinct. Since they belong to a generating system of
loops on $\Gamma$ we see that we may choose $\alpha_I$ as
generators as well. They obviously generate $\pi_1(\tilde{\Gamma})$.\\
$\Box$\\
The argument can be straightforwardly generalized to the case of a
multiloop $\alpha_1,..,\alpha_r$ by considering the pattern of surfaces
made by the union of those $r$ loops. In this way, the choices of open paths
$\rho_\Box$ between $p$ and the plaquettes $\Box$ made in (4.8) is
consistent because the $\Box$ appear in one and only one of the
non-overlapping, simple, homotopically trivial loops. As already mentioned
earlier, the Wilson action is not affected by such a choice.\\
Consider the case that the loop $\alpha$ contains a homotopically trivial
loop $\lambda=[\lambda\circ\gamma^{-1}]\circ\gamma$. The loop in the bracket
is homotopically trivial if $\gamma$ is in the same homotopy class as
$\lambda$. Thus we are able to express $\alpha$ in the manner
described above in terms of non-overlapping simple homotopically trivial
loops and our favourite homotopically non-trivial generators $\gamma$. \\
Summarizing, we have shown that
\ba
& & \chi(\alpha_1,..\alpha_r)=\frac{1}{Z}
\prod_\Box\int_G
d\mu_H(g_\Box)\exp(-\beta S_{Wilson}) \times\\
& & \left\{ \begin{array}{ll}
\prod_{i=1}^r \mbox{tr}("\prod_{\Box\in\alpha_i}g_\Box") &\mbox{: on }
R^2 \mbox{ and}\\ \int_G d\mu_H(g)
\prod_{i=1}^r
\mbox{tr}("\prod_{\gamma,\Box\in\alpha_i}g_\Box g") & \mbox{: on }
R^1\times S^1
 \end{array} \right. \nonumber
\ea
where the "" denote that one has to order the variables involved correctly
and that each of the variables could occur more than once and in particular
also its inverse. We have denoted by $g_\Box$ and $g$ plaquette and
homotopically non-trivial loop ($\gamma$) integration variables. Always
$\chi(0)=1$.

\subsection{The general form of the generating functional}

We now actually perform the lattice integration of any product of Wilson
loop functionals, thereby extending the results of \cite{25}.\\
Following the remarks at the end of section 2, we know the vacuum
expectation value of a multiloop functional when we know the vacuum
expectation value of its decomposition into loop networks. So, let us denote
the non-overlapping pieces of $\alpha_1,..,\alpha_r$ by
$\gamma_1,..,\gamma_m,\;m\ge r$ and the the holonomically non-trivial loop
by $\gamma$. Then, according to lemma 4.1, we can write the multi-loop
functional as a certain linear combination of the states
\be \label{4.1}
T_{\Gamma(a,N_x,N_y),(\vec{\pi},\pi),c}=\mbox{tr}[\otimes_{k=1}^m
\pi_k(h_{\gamma_k}(A))
\otimes \pi(h_\gamma(A))\cdot c] \;.
\ee
Let $Ar(\gamma_k)$ denote the area of the surface enclosed by $\gamma_k$.
According to lemma 4.2, each $\gamma_k$ can be expressed as a certain
product of the $n_k=Ar(\gamma_k)/a^2$ plaquette loops that are enclosed by
$\gamma_k$, each of them appearing once and only once, and, due to their
non-overlapping character, these sets of plaquette loops for the different
loops $\gamma_k$ are mutually different , that is, they provide independent
integration variables ! This is a special feature of two dimensions for
the plane and the cylinder.\\
So, let $\gamma_k=\beta_{k,1}\circ..\circ\beta_{k,n_k}$ then due to
$\pi_k(h_{\gamma_k})=\pi_k(h_{\beta_{k,1}})..\pi_k(h_{\beta_{k,n_k}})$
we end up doing the following basic integral ($\sigma$ an arbitrary
irreducible representation)
\be \label{4.3}
I_\sigma(\beta):=\int_G d\mu(g)
\sigma(g),\;d\mu(g)=\exp(-\beta[1-\frac{1}{N}\Re\mbox{tr}(g)]) d\mu_H(g) \;.
\ee
The measure $d\mu$ is conjugation invariant and therefore $I_\sigma$ has
to be
proportional to $\sigma(1)$ according to the lemma of Schur. By taking the
trace we see that
\be \label{4.4}
I_\sigma(\beta)=\sigma(1)
J_\sigma(\beta),\;J_\sigma(\beta):=\frac{1}{d_\sigma}\int_G
d\mu(g)\chi_\sigma(g) \ee
where $\chi_\sigma=\mbox{tr}\sigma$ is the character of the irreducible
representation and $d_\sigma$ is its dimension.\\
Since the measure for the loop $\gamma$ is just the Haar measure, the
vacuum expectation value of our loop-network is non-vanishing only if
$\pi$ is the trivial representation $0$. Therefore we get altogether
\be \label{4.5}
\chi(\Gamma(a,N_x,N_y),(\vec{\pi},\pi),c)=\delta_{\pi,0}\prod_{k=1}^m
[\frac{J_{\pi_k}(\beta)}{J_0(\beta)}]^{n_k} d_c \ee
where $d_c$ is the dimension of the irreducible subspace specified by $c$.
Here we have made use of the fact that $c$ is a projector and that
in the decomposition into irreducibles of $\otimes_{k=1}^n \pi_k$, the
irreducible
representation specified by c is contained.\\
Note that the expression (\ref{4.5}) is completely insensitive to the
size of the lattice due to the fact that the plaquettes are
non-interacting. Therefore the thermodynamic limit is already taken.
The task of taking the continuum limit now reduces to proving that
\be \label{4.6}
\omega(\sigma,\alpha):=\lim_{\beta\to\infty}
[\frac{J_\sigma(\beta)}{J_0(\beta)}]^{g_0^2 Ar(\alpha)\beta}
\ee
exists.\\
The proof goes as follows :\\
For $\beta\to\infty$, the integrand of $J_\sigma(\beta)$ is concentrated
at the identity, so it will be sufficient to calculate the integral
for $g$ in a neighbourhood $U$ of the identity. To that effect, write
$g=e^A\mbox{ where }A=t^I\tau_I\in L(G)$ is in the Lie algebra of $G$ and
$t^I$ are real parameters in a neighbourhood of zero. We thus have
upon inserting $g=1_N+A+\frac{1}{2}A^2+o(A^3)$
\be 1-\frac{1}{N}\Re\mbox{tr}(g)=-\frac{1}{2}\mbox{tr}(A^2)+o(A^3)
=\frac{1}{2}\sum_{I=1}^{\dim(G)}(t^I)^2+o(A^3) \ee
where the term of first order in $A$ vanishes because it is trace-free
($L(G)$ is semi-simple). We have
also used the normalization tr$(\tau_I\tau_J)=-N\delta_{IJ}$.\\
Similarly, we have an expansion for the $\sigma$th irreducible representation
of $G$ given by $\sigma(g)=\sigma(1)+X+\frac{1}{2}X^2+o(X^3)
\mbox{ where }
X=t^I X_I$ is the representation of the Lie algebra element $A$
in the $\sigma$-th irreducible representation. Then we have
\be \chi_\sigma(g)=d_\sigma+t^I\mbox{tr}(X_I)+\frac{1}{2}
t^I t^J\mbox{tr}(X_I X_J) +o(X^3) \;.\ee
According to the Baker-Campbell-Hausdorff formula \cite{8} we have that
\be e^{t^I\tau_I}e^{s^I\tau_I}=e^{r^I(s,t)\tau_I},\mbox{ where }
r^I(s,t)=s^I+t^I-\frac{1}{2}f^I\;_{JK}s^J
t^K+o(s^2,t^2,s^3,s^2t,st^2,t^3) \ee
and where $f^I_{JK}$ are the structure constants of the semi-simple Lie
algebra of $L(G)$ which therefore are completely skew. Now, the Haar
measure can be written \cite{8}
\be d\mu_H(e^{t^I\tau_I})=\frac{d^{\dim(G)}t}{\det(\frac{\partial r^I(s,t)}
{\partial s^J})_{s=0}}=\frac{d^{\dim(G)}t}{1+o(t^2)} \ee
since $\det(\partial r/\partial s)_{s=0}=\det(1+\frac{1}{2} t^I R_I
+o(t^2))=1+\frac{1}{2}\mbox{tr}
(t^I R_I)+o(t^2)=1+o(t^2)\mbox{ where }(R_I)^J_K=f^J\;_{IK}$ is the
I-th basis vector of $L(G)$ in the adjoint
representation which is trace-free.\\
We now change coordinates $t\to\sqrt{\beta}t$, insert (4.15), (4.16) and
(4.18)
into (4.14), write an expansion in $1/\sqrt{\beta}$ and integrate with the
result \be
J_\sigma(\beta)-J_0(\beta)=\frac{1}{d_\sigma\beta}J_0(\beta)
\frac{1}{2}\mbox{tr}(\sum_{I=1}^{\dim(G)} (X^I)^2)+o(1/\beta^2) \;.\ee
But $\sum_I (X_I)^2=-\lambda_\pi \pi(1)$ is the Casimir
invariant and $\lambda_\pi$ its eigenvalue. Therefore we arrive
finally at $\omega(\pi,\gamma)=\exp(-\frac{1}{2}\lambda_\pi g_0^2
Ar(\gamma)$ and thus
\be
\chi(\Gamma[\{\gamma_k\},\gamma\},(\vec{\pi},\pi),c)=\delta_{\pi,0}
d_c e^{-\frac{1}{2}g_0^2\sum_{k=1}^m \lambda_k Ar(\gamma_k)} \;. \ee

\section{Verification of the axioms and comparison with the Hamiltonian
formalism}

Let us first verify the axioms.\\
III) The generating functional (4.20)
clearly depends only on the areas of the various loops involved
and therefore is not only invariant under the Euclidean group (rather, the
symmetry group of the metric on the cylinder) but even
under area-preserving diffeomorphisms. \\
IV) Reflection positivity is also satisfied because after dividing by the
space $\cal N$ of null vectors in V we
obtain a scalar product which is positive definite as we will show now
by employing the algorithm displayed in \cite{11} :\\
Consider a multiloop $\{\alpha_1,..,\alpha_s\},\;s\le r$ which is composed,
among others, of homotopically non-trivial loops.
Let $\gamma$ be the horizontal loop at $t=0$ and write every
homotopically non-trivial loop $\eta$ occurring in the multiloop
$\{\alpha_1,.., \alpha_s\}$ as $\eta=[\eta\circ\gamma^{-1}]\circ\gamma$
where the loop in brackets is homotopically trivial, thereby obtaining
a multiloop $\tilde{\alpha}_1,..,\tilde{\alpha}_s$ whose homotopically
trivial contribution comes from $\gamma$ only. It follows that all
the vectors in V can be written as linear combinations of loop networks
where the homotopically non-trivial contribution comes form $\gamma$.
The special feature of the loop $\gamma$ is that it is left
invariant under the time reflection operation. We now write
\[
T_{\Gamma,(\vec{\pi},\pi),c}=T_{\Gamma-\gamma,\vec{\pi},c^j_i}\pi^i_j(h_\gamma)
\]
where $c^j_i$ is the matrix obtained from $c$ by fixing the last two indices
of both its $m+1$ fold multi-indices to be $i,j$. Now, observing that
for $\Gamma,\Gamma'$ supported in the positive time half space it is true
that $\Theta(\Gamma-\gamma),\Gamma'-\gamma$ contain only topologically
trivial
loops enclosing disjoint areas in the two-dimensional spacetime. The
non-interactive nature of the measure therefore implies, using the basic
integral $\int_G d\mu_H \bar{\pi}^i_j \pi^{'k}_l=1/d_\pi \delta_{\pi,\pi'}
\delta^{ik}\delta_{jl}$, that
\be
<\Theta T_{\Gamma,(\vec{\pi},\pi),c},T_{\Gamma',(\vec{\pi}',\pi'),c'}>
=\overline{\chi(\Theta(\Gamma-\gamma),\vec{\pi},c^i_j) }
\chi(\Gamma'-\gamma,\vec{\pi}',c^{'i}_j)\frac{1}{d_\pi}\delta_{\pi,\pi'}
\ee
According to (4.20) each of the characteristic functionals on the right
hand side of (5.1) are proportional to $\pi^i_j(1)$ and the usual trace
argument
shows, using the fact that $\chi$ is in particular invariant under
$\Theta$, that
\be
T_{\Gamma,(\vec{\pi},\pi),c}-\frac{\chi(\Gamma-\gamma,\vec{\pi},c^i_i)}
{\sqrt{d_\pi}}
\chi_\pi(h_\gamma) \ee
is a null vector. Therefore the physical Hilbert space is just given
by ${\cal H}:=\overline{V/{\cal N}}=L_2(G,d\tilde{\mu}_H),\;\tilde{\mu}_H$
being the effective measure obtained from $\mu_H$ by restricting
integration to gauge invariant functions (characters, that is, functions
on the Cartan subgroup of $G$), which thus leaves us with a
positive definite sesquilinear form.\\
V) We prove ergodicity as follows :\\
$\gamma(t):=T(t)\gamma$ is the horizontal loop at time t. Now
let $\alpha(t):=\gamma(t)\circ\gamma^{-1}$, then we have by the
representation property
\be \chi_\pi(h_{\gamma(t)})=\mbox{tr}(\pi(h_{\alpha(t)})
\pi(h_\gamma)) \ee
so that with respect to ( , ) we have
\be
\chi_\pi(h_{\gamma(t)})
=\frac{1}{d_\pi}[\int_{\AmGb} d\mu
\chi_\pi(\alpha(t))]\chi_{\pi}(h_\gamma)=e^{-\frac{1}{2}g_0^2\lambda_\pi
L_x t}\chi_\pi(h_\gamma) \;. \ee
Therefore
$ \lim_{t\to\infty}\frac{1}{t}\int_0^t ds T(s) \chi_\pi=
\delta_{0,\pi}=\int d\mu \chi_\pi $ and the proof is complete.
Meanwhile we see from the definition of the Hamiltonian as the generator
of time translations that
\[
(\chi_{\pi'},T(t)\chi_\pi)=\exp(-\frac{1}{2}\lambda_\pi
g_0^2 L_x t)\delta_{\pi,\pi'}\stackrel{!}{=}(\chi_{\pi'},
\exp(-t H)\chi_\pi) \]
and the completeness of the $\chi_\pi$ on $L_2(G,d\tilde{\mu}_H)$ allows
us to conclude that \beq H=-\frac{g_0^2}{2}L_x\Delta \eeq
is the coordinate representation for the Hamiltonian where $\Delta$ is
the Casimir operator on $G$. The unique vacuum
vector is $\Omega=1$, the only vector annihilated by the Hamiltonian.\\
We therefore have a simple finite dimensional model in front of us for which
the proposed axioms are indeed verified, thus proving non-triviality of
the axioms.\\
\\
Let us now make contact with the Hamiltonian analysis :\\
After splitting 2-dimensional Minkowski spacetime into space and time, the
action for 2-dimensional YM theory becomes \cite{23}
\beq S=\int_R dt\int_\Sigma dx[\dot{A}_I E^I-[-\Lambda^I{\cal G}_I
+\frac{g_0^2}{2} E^I E^I]]
\eeq
where $\Sigma=\Rl\mbox{ or }S^1$. Here $A=A_x$ is the (pull-back to $\Sigma$
of the) G connection and $E=\frac{1}{g_0^2}(\partial_t A_x-\partial_x A_t
+[A_t,A_x])$ is its electric field.  The Gauss constraint and the
Hamiltonian are respectively given by (a prime means a derivative with
respect to x) \beq
{\cal G}_I=E_I'+[A,E]_I \mbox{ and } H=\int_\Sigma dx
\frac{g_0^2}{2} E^I E^I. \ee
Now, multiplying the Gauss constraint with $E^I$ we infer that
the Hamiltonian density is a constant on the constraint surface
$\frac{1}{2}(E^I E^I)'=0$
which immediately implies that the energy is infinite on the plane unless
that constant is zero. This in turn implies $E^I=0$ and the theory
becomes trivial on the plane.\\
On the cylinder, however, the theory is less trivial, the Hamiltonian is
just given by  $H=\frac{g_0^2 L_x}{2}(E^I E^I)$
which remains finite for finite $E$ due to the compactness of the x
direction.\\
We now quantize this theory along the Dirac approach, that is, we solve the
constraints by imposing it on the states which we choose in the connection
representation.\\
The canonical commutation relations resulting from the Poisson bracket\\
$\{A_I(x),E^J(y)\}=\delta_I^J \delta(x,y)$ for the canonical pair
$A_I, E^I$ are met if we choose the following operator representation
\beq (\hat{A}_I(x)\psi)(A)=A_I(x)\psi(A),\;(\hat{E}_I(x)\psi)(A)=-i
\frac{\delta}{\delta A_I(x)}\psi(A)\;. \eeq
Imposing the Gauss constraint on the state space then immediately tells
us that they have to be gauge invariant, that is, they have to built from
Wilson loops. But the only loop that we have is $\gamma=\Sigma=S^1$ itself,
therefore a complete set of physical states is given by the characters
$\chi_\pi(h_\gamma(A))$ which form an orthonormal base on
$L_2(G,d\tilde{\mu}_H)$.\\
Direct evaluation reveals ($g=h_\gamma(A)$)
\be \hat{H}(x)\chi_\pi(g)=\frac{g_0^2 L_x}{2}X_I(g) X_I(g) \chi_\pi(g)
=\frac{g_0^2 L_x}{2}\Delta(g)\chi_\pi(g)=-\frac{g_0^2 L_x\lambda_\pi}{2}
\chi_\pi(g) \ee
where $X_I(g)=\mbox{tr}(g\tau_I\partial/\partial g)$ is the left invariant
vector field on $G$.
This demonstrates exact agreement between both approaches.
\\
\\
\\
{\large Acknowledgements}\\
\\
I am grateful for many important insights obtained in the course of
discussions with Abhay Ashtekar, Jurek Lewandowski, Donald
Marolf and Jos\'e Mour\~ao.\\
The author was supported in part by the NSF Grant PHY93-96246 and the Eberly
research fund of The Pennsylvania State University.


\begin{thebibliography}{99}

\parskip=-7pt

\bibitem[1]{15} A.\ Ashtekar, Phys.\ Rev.\ {\bf D36} (1987)1587
\bibitem[2]{5}  R.\ Giles, Phys.\ Rev. {\bf D8} (1981) 2160
\bibitem[3]{24} T.S. Wu, C.N. Yang, Phys. Rev. {\bf D12} (1975) 3843
\bibitem[4]{2}  A.\ Ashtekar, J.\ Lewandowski, {\bf Representation Theory of
                analytic holonomy $C^*$ algebras}, in ``Knots
                and quantum gravity'', J.\ Baez (ed), Oxford University
                Press, 1994
\bibitem[5]{3}  A.\ Ashtekar, C.\ J.\ Isham, Class. Quantum Grav., {\bf 9}
                (1992) 1069
\bibitem[6]{4}  D.\ Marolf, J.\ M.\ Mour\~ao, {\bf On the support of the
                Ashtekar-Lewandowski measure}, accepted for publication by
                {\bf Comm. Math. Phys.}
\bibitem[7]{26} A.\ Ashtekar, J.\ Lewandowski, D.\ Marolf, J.\ Mour\~ao,
                T.\ Thiemann, {\bf A manifestly gauge invariant approach to
                quantum theories of gauge fields}, in : Geometry of
                Constrained Dynamical Systems, J. Charap (editor), Cambridge
                University Press, Cambridge, 1994
\bibitem[8]{13} E.\ Seiler, {\bf Gauge theories as a problem of constructive
                quantum field theory and statistical mechanics}, Lecture
                notes in Physics, v. 159, Springer-Verlag, Berlin,
                Heidelberg, New York, 1982
\bibitem[9]{11} J.\ Glimm, A.\ Jaffe, {\bf Quantum Physics}, 2nd ed.,
                 Springer-Verlag, New York (1987)
\bibitem[10]{20} C. Rovelli, L. Smolin, ``Spin-networks and quantum
                 gravity", Preprint CGPG-95/4-4\\
                 J. Baez, ``Spin network states in gauge theory", Adv.
                 Math. (in press)
\bibitem[11]{21} T. Thiemann, ``The inverse loop transform", Preprint
                 CGPG-95/7-2
\bibitem[12]{12} Y.\ Yamasaki, {\bf Measures on infinite dimensional spaces},
                World Scientific, Philadelphia (1985)
\bibitem[13]{22} T. Thiemann, ``A Minlos theorem for gauge theories",
                 (in preparation)
\bibitem[14]{8} M.\ Creutz, {\bf Quarks, Gluons and Lattices}, Cambridge
                University Press, New York (1983)
\bibitem[15]{19} V. Rivasseau, {\bf From perturbative to constructive
                 renormalization}, Princeton University Press, Princeton,
                 1991
\bibitem[16]{23} A. Ashtekar, J. Lewandowski, D. Marolf, J. Mour\~ao, T.
                 Thiemann, ``Euclidean Yang-Mills Theory in two
                 dimensions : A complete solution", Preprint CGPG-95/7-3
\bibitem[17]{25} L. Gross, C. King, A. Sengupta, Ann. Phys. {\bf 194} (1989)
                 65-112\\
                 S.\ Klimek, W.\ Kondracki, Comm.\ Math.\ Phys.\ {\bf 113}
                 (1987) 389-402\\
                 V.\ A.\ Kazakov, Nucl.\ Phys.\ {\bf B79} (1981) 283
\bibitem[18]{16} S.\ Helgason, {\bf Differential Geometry, Lie Groups and
                 Symmetric Spaces}, Academic Press, San Diego, 1978


\end{thebibliography}
\end{document}